\documentstyle[preprint,aps,epsfig]{revtex}

\begin{document}
\draft
\preprint{
\vbox{ \hbox{SNUTP\hspace*{.2em}$99-022$}}
}

\title{Large Extra Dimension Effects \\
on the Spin Configuration of the Top Quark Pair \\
at $e^+ e^-$ Colliders}

\author{
$^{(a)}$Kang Young Lee,
$^{(a,b)}$H.S. Song,
$^{(a)}$JeongHyeon Song,
and $^{(b)}$Chaehyun Yu.
}

\address{
$^{(a)}$ Center for Theoretical Physics,
     Seoul National University, Seoul 151-742, Korea \\
$^{(b)}$ Department of Physics,
Seoul National University, Seoul 151-742, Korea
}

\maketitle
\vspace{2cm}

\begin{abstract}
Large extra dimension effects on the spin configuration of the top 
quark pair at the $e^+ e^-\to t\bar{t}$ process are studied. It is 
shown that the TeV scale quantum gravity effects cause significant 
deviations from the Standard Model predictions for the spin 
configuration in the off-diagonal basis: they lead to substantial 
cross sections of the like-spin states of the top quark pair, which 
vanish in the SM; they weaken the pure dominance of the processes, 
the Up-Down (Down-Up) spin states for the left-handed (right-handed) 
beam. In addition it is shown that the angular cut $-0.5<\cos\theta<0$ 
is very effective to determine the sign of the quantum gravity corrections.
\end{abstract}

%\tightenlines
%%%%%%%%%%%%%%%%%%%%%%%%%%%%%%%%%%%%%%%%%%%%%%%%%%
%\section{Introduction}
%%%%%%%%%%%%%%%%%%%%%%%%%%%%%%%%%%%%%%%%%%%%%%%%%%

There has been an increasing interest 
in the low scale quantum gravity 
as an extension of the Kaluza--Klein (KK) scenario.
Recently Arkani-Hamed, Dimopoulos, and Dvali (ADD) \cite{ADD1}
have suggested that the size of the extra dimensions
could be large enough to be detectable
if we confine the matter fields 
to the 4-dimensional world where we live.
According to their idea, 
the weakness of the gravity in our world
is caused by the suppression factor from the
large extra dimensions since gravitons are the only fields
freely propagating in the whole $(4+N)$-dimensional spacetime.
Considering the macroscopic Gauss' Law for the Newtonian gravity,
ADD have derived the relation between the Planck scale and the
size of the extra dimensions as
\begin{equation}
M_{Pl}^2 \sim M_S^{N+2} R^N,
\end{equation}
where $M_S$ is the only fundamental scale of nature,
of which the value is comparable with the electroweak scale.
The $N=1$ case is excluded by this simple relation
because the corresponding size of the extra dimension 
is order of $10^{13}$ cm 
if the $M_S$ is at a few TeV order 
which is phenomenologically interesting.
The $N=2$ case which implies mm scale extra dimensions
is not excluded by the current macroscopic measurement of
gravitational force \cite{macro}.
For the cases of $N>2$ there exist
no other serious constraints up to now.
The hierarchy of the Planck scale $M_{Pl}$ 
and the electroweak scale $M_{_W}$ is reduced to
the revelation of the effects of the large extra dimensions.

Indeed interesting is that this idea is testable
by the collider phenomenologies in the near future.
When the graviton momentum does not exceed the scale of $M_S$,
the spacetime where collisions take place 
can be approximately described
by the linear expansion around the flat metric.
Within the framework of the linearized gravity,
the effective action in the 4-dimensional spacetime
is derived after compactifying the extra dimensions,
which leads to corresponding Feynman rules\cite{han,giudice}.
The KK reduction 
from the whole $(4+N)$-dimension to our 4-dimension
yields towers of massive KK states
in the 4-dimensional effective theory,
of which the massive spectrum is cutoff at the scale $M_S$. 
Each graviton in such KK towers
interacts with the ordinary matter fields 
with the couplings suppressed by the Planck scale.
The production of a single graviton is enhanced by the kinematic
factor and has been studied as a source of the missing energy 
in the $e^+ e^- \to \gamma E\hspace{-3.5mm}/$ or $p \bar{p} 
\to \gamma E\hspace{-3.5mm}/$ process
\cite{giudice,Peskin}.
The indirect effects of massive graviton exchange
may be enhanced by the sum of the tower of the KK states
and provide various signals in the collider
phenomenologies \cite{Hewett,Rizzo,top,Agashe,Balazs,Kingman,lee}.
In particular the spin two nature of the gravitons
will result in some characteristic effects on the polarization observables
\cite{lee}.

In the present work, we consider the top quark pair production
in the Next $e^+ e^-$ Linear Collider (NLC) \cite{LC}.
It is a promising testing ground for new physics effects
because of the heavy mass of top quark.
The presence of large extra dimensions affects
the top quark pair production at $e^+e^-$ collisions through
the interactions between the fermions and 
the Kaluza-Klein tower of gravitons.
In the view point of our 4-dimensional world,
the massless gravitons freely propagating in the $(4+N)$-dimensional bulk
are massive spin two, spin one and spin zero gravitons under the
compactification of the extra dimensional manifold.
Since the remarkable successes of the SM in explaining all
the high energy experiments can be retained by assuming the SM
fields confined on our world, {\it i.e.}
$T_{\mu j}=T_{ij}=0$ $(\mu=1, \cdots 4; i,j=5, \cdots  (4+N))$,
only the spin two gravitons and the dilaton modes of the
spin zero gravitons interact with the ordinary matter.
In addition the coupling of the dilaton modes to the electron
is to be neglected at high energy collisions, since it is
proportional to the fermion mass.
Thus the process $e^+e^- \to t\bar{t}$ receives the
effects of TeV scale quantum gravity by the $s$-channel Feynman
diagram mediated by the spin two Kaluza-Klein gravitons.
Furthermore, the produced top quark is known to be
in the unique spin configuration at the polarized $e^+ e^-$ 
collider\cite{parke1}
and the information of the top spin 
is not lost through hadronization
since its lifetime is too short for the top quark to constitute hadrons.
We can read out the information on the top polarization 
through the angular distribution of the decay 
products\cite{spin-correlation}.
The neutral current nature of the graviton interactions
leaves the electroweak decay of top quark pair intact,
implying that the SM prediction of the
angular correlations between the decay products and
the spin orientation of each top quark 
remains valid\cite{spin-correlation}.
Therefore the spin configuration of the top quark pair
can be a probe of the effects of KK gravitons.

For the process
\begin{equation}
e^-(k_1) + e^+(k_2) \rightarrow t(p_1) +
\overline{t}(p_2)
\;,
\end{equation}
the scattering amplitude of the $s$-channel Feynman diagram
mediated by the spin two gravitons summed over the Kaluza-Klein tower
can be written by
\begin{eqnarray}
\label{top-amp}
{\cal M}_G = \frac{\lambda}{M_S^4}&\Big[& (k_1-k_2)\cdot(p_1-p_2)
   \overline{v}(k_2) \gamma_\mu u(k_1) \overline{u}(p_1)\gamma^\mu v(p_2)
    \nonumber \\
     & +& \overline{v}(k_2)(p_1\hspace{-3.5mm}/ - p_2 \hspace{-3.5mm}/ \hspace{2mm})
      u(k_1) \overline{u}(p_1)(k_1\hspace{-4mm}/ - k_2\hspace{-4mm}/ 
      \hspace{2mm}) v(p_2) \hspace{2mm}\Big],
\end{eqnarray}
where the order one parameter $\lambda$ 
depends on the number of extra dimensions and the compactification models.
Since the exact value of the $\lambda$ including sign is not determined
unless the full quantum gravity theory is provided,
two representative cases of $\lambda=\pm$ are considered
in the following discussion\cite{Hewett}.
It is also to be noted that the amplitude in Eq.(\ref{top-amp})
as well as the SM amplitudes at the tree level
are CP invariant.

In order to analyze the spin configuration of the top quark pair,
let us briefly review a generic spin basis discussed in Ref.\cite{parke1}.
We define the spin states of the top quark and top anti-quark
in their own rest-frame by decomposing their spins
along the reference axes $\hat{\eta}$ and $\hat{\bar{\eta}}$,
respectively.
The CP invariance,
which is valid at the tree level even with the large extra dimension
effects,
does not allow the T odd quantity, $i.e.$
$\vec{\sigma}_t \cdot (\vec{k_1}|_{(t{\rm -rest})} \times
\vec{p_2}|_{(t{\rm -rest})})$
where the $\vec{\sigma}_t$ is the spin of the top quark,
and $\vec{k_1}|_{t{\rm -rest}}$ and $\vec{p_2}|_{t{\rm -rest}}$
are the momenta of the electron and the top anti-quark
in the rest frame of the top\cite{parke1,Kane}.
Thus the top and anti-top spins are to lie in the production plane.
The spin four-vectors of the top quark pair are chosen to be
back-to-back in the zero momentum frame.
The $\hat{\eta}$ is
expressed by an angle $\xi$, the angle between $\hat{\eta}$ and the
top anti-quark momentum in the rest frame of the top quark.
The usual helicity basis is obtained by taking $\xi=\pi$.

In this general spin basis the differential cross sections 
of the $e^+ e^- \to t\bar{t}$ process 
with the large extra dimension effects are
\begin{eqnarray}
\displaystyle 
\frac{d\sigma}{d\cos\theta}
  (e^-_Le^+_R\rightarrow t_\uparrow \bar{t}_\uparrow ~{\rm or}~
  t_\downarrow \bar{t}_\downarrow) &=& 
   \frac{N_c\pi\alpha^2\beta}{2s}  
  |\tilde{A}_{L}\cos\xi-\tilde{B}_{L}\sin\xi|^2 ,\nonumber \\
\frac{d\sigma}{d\cos\theta}
  (e^-_Le^+_R\rightarrow t_\uparrow \bar{t}_\downarrow ~{\rm or}~
  t_\downarrow \bar{t}_\uparrow) &=& 
   \frac{N_c\pi\alpha^2\beta}{2s} 
  |\tilde{A}_{L}\sin\xi+\tilde{B}_{L}\cos\xi\pm\tilde{D}_{L}|^2,\nonumber \\
\frac{d\sigma}{d\cos\theta}
  (e^-_Re^+_L\rightarrow t_\uparrow \bar{t}_\uparrow ~{\rm or}~
  t_\downarrow \bar{t}_\downarrow) &=& 
   \frac{N_c\pi\alpha^2\beta}{2s} 
  |\tilde{A}_{R}\cos\xi-\tilde{B}_{R}\sin\xi|^2 ,\nonumber\\
\label{general}
\frac{d\sigma}{d\cos\theta}
  (e^-_Re^+_L\rightarrow t_\uparrow \bar{t}_\downarrow ~{\rm or}~
  t_\downarrow \bar{t}_\uparrow) &=& 
   \frac{N_c\pi\alpha^2\beta}{2s} 
  |\tilde{A}_{R}\sin\xi+\tilde{B}_{R}\cos\xi\mp\tilde{D}_{R}|^2, 
\end{eqnarray}
where $t_\uparrow\;(t_\downarrow)$
denotes the top spin along (against) the $\hat{\eta}$, 
$N_c$ is the number of color, $\alpha$ is the fine structure
constant, $\beta=\sqrt{1-{4m_t^2}/{s}}$ and
\begin{eqnarray}
\tilde{A}_{L} &=& \frac{1}{2}(f_{LL}+f_{LR})
       		\sin\theta\sqrt{1-\beta^2}-f_G\sin 2\theta
		\sqrt{1-\beta^2} ,\nonumber \\
\tilde{B}_{L} &=& \frac{1}{2}\bigg[f_{LL}(\cos\theta+\beta)
       		+f_{LR}(\cos\theta-\beta)\bigg]
		-f_G\cos2\theta
		,\nonumber \\
\tilde{D}_{L} &=& \frac{1}{2}\bigg[f_{LL}(1+\beta\cos\theta)
       +f_{LR}(1-\beta\cos\theta)\bigg] -f_G\cos\theta
	,\nonumber \\
\tilde{A}_{R} &=& \frac{1}{2}(f_{RR}+f_{RL})\sin\theta\sqrt{1-\beta^2}
		-f_G \sin 2\theta\sqrt{1-\beta^2} ,\nonumber \\
\tilde{B}_{R} &=& \frac{1}{2}\bigg[f_{RR}(\cos\theta+\beta)
       +f_{RL}(\cos\theta-\beta)\bigg]
	-{f_G}\cos2\theta
                ,\nonumber \\
\tilde{D}_{R} &=& \frac{1}{2}\bigg[f_{RR}(1+\beta\cos\theta)
       +f_{RL}(1-\beta\cos\theta) \bigg]
	-{f_G}\cos\theta
.
\end{eqnarray}
The large extra dimension
effects are altogether included in the quantity $f_G$ defined by
\begin{equation}
\label{fG}
f_G = \frac{\beta s^2}{4\alpha} \frac{\lambda}{M_S^4}
\;.
\end{equation}
Here $f_{IJ}$'s $(I,J=L~ {\rm or}~ R)$ are
\begin{equation}
f_{IJ} = Q_\gamma(e)Q_\gamma(t)+Q_Z^I(e)Q_Z^J(t)\frac{1}{\sin^2\theta_W}
      \frac{s}{s-M_Z^2},
\end{equation}
and $\theta$ is the scattering angle of the top quark
with respect to the electron beam.
The electric charges and couplings to the $Z$ boson
of the electron and the top quark are given by
\begin{eqnarray}
Q_\gamma(e)=-1\hspace{18.7mm},&&\hspace{10mm}Q_\gamma(t)=\frac{2}{3},\nonumber\\
Q_Z^L(e)=\frac{2\sin^2\theta_W-1}{2\cos\theta_W},
  &&\hspace{10mm}Q_Z^R(e)=\frac{\sin^2\theta_W}{\cos\theta_W},\nonumber\\
Q_Z^L(t)=\frac{3-4\sin^2\theta_W}{6\cos\theta_W},
  &&\hspace{10mm}Q_Z^R(t)=-\frac{2\sin^2\theta_W}{3\cos\theta_W}.
\end{eqnarray}

A comment on the behaviors of the $f_{IJ}$'s are in order here
for they play a central role in the SM background.
Numerical results show that
the $f_{IJ}$'s are negative quantity
and their dependence on the beam energy
is weak at high energy.
For instance, their numerical values at $\sqrt{s}=$500 GeV 
are
\begin{equation}
f_{LL}=-1.21, \quad f_{LR}=-0.43,\quad
f_{RL}=-0.20, \quad f_{RR}=-0.87,
\end{equation}
indicating $f_{LL} > f_{LR}$ and $f_{RR} > f_{RL}$.

There exist the angles $\xi_{L}$ and $\xi_{R}$ such that the
differential cross sections for the
$t_\uparrow \bar{t}_\uparrow$ and $t_\downarrow \bar{t}_\downarrow$,
{\it i.e.,} like-spin states
vanish for the left- and right-handed electron beam, respectively.
It is called the ``off-diagonal basis",
of which the name originated in this feature\cite{parke1}.
This is general in any $2\rightarrow 2$ process 
if the CP is conserved
and the spin four-vectors of the outgoing particles 
are back-to-back in the zero momentum frame.

In the SM, the angles  $\xi_{L,R}$ are taken to be
\begin{equation}
\label{xi}
\cos\xi_I = -\frac{B_I}{\sqrt{A_I^2+B_I^2}}\;,
\quad
\sin\xi_I = -\frac{A_I}{\sqrt{A_I^2+B_I^2}}\;,
\end{equation}
where $I=L,R$, $A_I=\tilde{A}_I|_{f_G=0}$, and $B_I=\tilde{B}_I|_{f_G=0}$.
The differential cross sections in Eq.(\ref{general}) are reduced to, 
in this SM off-diagonal basis,
\begin{eqnarray}
\displaystyle
\frac{d\sigma}{d\cos\theta}
  (e^-_Le^+_R\rightarrow t_\uparrow \bar{t}_\uparrow ~{\rm or}~
   t_\downarrow \bar{t}_\downarrow) &=&
    \frac{N_c\pi\alpha^2\beta}{2s} f_G^2
   \Bigg[ \sin\xi_L\cos2\theta-\cos\xi_L\sin2\theta \sqrt{1-\beta^2}
    \Bigg]^2,\nonumber \\
\frac{d\sigma}{d\cos\theta}
  (e^-_Le^+_R\rightarrow t_\uparrow \bar{t}_\downarrow ~{\rm or}~
   t_\downarrow \bar{t}_\uparrow) &=&
     \frac{N_c\pi\alpha^2\beta}{2s}
   \Bigg[  \sqrt{ A_{L}^2+ B_{L}^2} \mp {D}_{L} \nonumber \\
&&  +{f_G}( \cos\xi_L \cos2\theta 
+ \sin\xi_L \sin2\theta \sqrt{1-\beta^2} 
\pm \cos\theta ) \Bigg]^2,\nonumber \\
\frac{d\sigma}{d\cos\theta}
  (e^-_Re^+_L\rightarrow t_\uparrow \bar{t}_\uparrow ~{\rm or}~
   t_\downarrow \bar{t}_\downarrow) &=&
    \frac{N_c\pi\alpha^2\beta}{2s} f_G^2
   \Bigg[ \sin\xi_R\cos2\theta-\cos\xi_R\sin2\theta \sqrt{1-\beta^2}
    \Bigg]^2,\nonumber \\
\frac{d\sigma}{d\cos\theta}
  (e^-_Re^+_L\rightarrow t_\uparrow \bar{t}_\downarrow ~{\rm or}~
   t_\downarrow \bar{t}_\uparrow) &=&
     \frac{N_c\pi\alpha^2\beta}{2s}
   \Bigg[  \sqrt{ A_{R}^2+ B_{R}^2} \pm {D}_{R} \nonumber \\
\label{cs-off}
&&  +f_G( \cos\xi_R \cos2\theta
+ \sin\xi_R \sin2\theta \sqrt{1-\beta^2}
\mp \cos\theta ) \Bigg]^2,
\end{eqnarray}
where $D_I=\tilde{D}_I|_{f_G=0}$.
				
There are two important characteristic features of the SM 
background.
First, the differential cross sections
for the like-spin states of the top quark pair vanish. 
Or we have chosen the spin configuration in that way.
Secondly, the process for $t_\uparrow \bar{t}_\downarrow$
$(  t_\downarrow \bar{t}_\uparrow)$ is dominant when the
left-handed (right-handed) electron beam is used.
This can be easily seen in Eq.(\ref{cs-off})
since the negative quantities $A$'s, $B$'s and $D$'s
are the same order of magnitude.
At high energy, the degree of this dominance is close to
100\%\cite{Why}.
This pure dominance of the Up-Down state for the left-handed
electron beam and the Down-Up state for the right-handed one
is fairly stable by the one-loop QCD corrections where 
the soft gluon emissions are dominant so that the QCD corrections
are factored out.
For instance, the ${\mathcal O}(\alpha_s)$ corrections
to the ratio of the cross section for
the dominant process to the total cross section
is just $\sim 0.01$\% while those to the total cross section 
is about $\sim 30$\%
at $\sqrt{s} \sim 500$ GeV\cite{Why,parke2}.

The TeV scale quantum gravity modifies these two features.
First the differential cross sections of the like-spin states
acquire the quantum gravity effects.
Secondly the presence of $f_G$ in the differential cross sections of the
dominant spin configurations corresponding to the incident beam
polarizations pollutes their pure dominance.
In Fig.~1 and 2 we plot the differential cross sections with respect to the
top quark scattering angle,
broken down to the spin configuration of the top quark pair
at $\sqrt{s}=500$ GeV with the left- and right-handed beams, respectively.
The solid lines denote the SM background,
and the dotted (dashed) lines include the
quantum gravity effects with $\lambda=+1$ $(\lambda=-1)$.
The $M_S$ is set to be $2.5$ TeV.
Figures 3 and 4 illustrate the same observables at $\sqrt{s}=1$ TeV.
At $\sqrt{s}=500$ GeV, the corrections 
are small over all the spin configurations, 
and become a little larger in the case of the right-handed electron beam.
Figures~3 and 4  show that the quantum gravity corrections 
increase with the beam energy.
In particular, the like-spin states,
which are zero in the tree level SM, gain sizable cross sections.

We present the change of the degree of dominance
corresponding to the beam polarization due to the large extra dimensions.
Table~1 shows the ratios of the cross section for the dominant process
to the total cross section 
at $\sqrt{s}=500$ GeV and $\sqrt{s}=1$ TeV for $\lambda=\pm 1$, 
which are expected to be stable by the one-loop QCD corrections.
As the beam energy increases enough,
the $\lambda=-1$ case gives rise to more deviations from the SM prediction.
And it can be easily seen that
the use of the high energy right-handed electron beam is more efficient
in detecting the corrections.

\vskip 1.0cm
\noindent
\begin{center}
\begin{tabular}{|c|cc|ccc|}\hline
 &\multicolumn{2}{|c|}{~~$\sqrt{s}=0.5$ TeV}
&\multicolumn{3}{|c|}{~~$\sqrt{s}=1$ TeV}
\\ \hline
{} &~ SM~ &~ SM $+$ QG~ & ~SM~ &~ SM $+$ QG~ &
~ SM $+$ QG~ \\ 
  &  &$(\lambda=+1)$ & & $(\lambda=+1)$ & $(\lambda=-1)$ \\ \hline
~${\sigma(e^-_Le^+_R\rightarrow t_\uparrow \bar{t}_\downarrow)}/
{\sigma(e^-_Le^+_R\rightarrow t \bar{t})}  $ ~
& ~99.26\% & 99.30\% & ~96.52\% & 86.62\% & 78.52\%\\
~${\sigma(e^-_Re^+_L\rightarrow t_\downarrow \bar{t}_\uparrow)}/
{\sigma(e^-_Re^+_L\rightarrow t \bar{t})}  $ ~
& ~99.56\% & 99.58\% & ~97.93\% & 82.48\% & 74.78\%\\
\hline
\end{tabular}
\end{center}
\vspace{0.5cm}
{Table~1}. {\it The ratios of the cross section of the dominant process
to the total cross section with $M_S=2.5$ TeV.}
\vskip 0.5cm

Finally let us observe that the angular distribution of the
cross sections have provided valuable information
on the nature of the interactions between gravitons and fermions.
According to the sign of the $\lambda$,
the quantum gravity corrections act in a different way.
In the dominant processes (the Up-Down spin state with
the left-handed electron beam and the Down-Up spin state with the
right-handed one),
the virtual graviton exchanges cause destructive (constructive)
interference with the SM diagrams to the backward direction
when $\lambda=+1$ $(\lambda=-1)$.
To the forward direction,
on the contrary,
constructive (destructive) interference occurs
when $\lambda=+1$ $(\lambda=-1)$.
In the next dominant processes
(the Down-Up spin state with $e^-_L$
and the Up-Down one with $e^-_R$),
enormous quantum corrections with $\lambda=+1$ 
to the backward direction are reduced and dispersed.
Therefore we suggest that angular cut, {\it e.g.,}
$-0.5 < \cos\theta<0$
be very effective to probe
the TeV scale quantum gravity corrections.
The usefulness of the angular cut is demonstrated in Table 2
through the ratios between the cross sections 
for the next-dominant and dominant processes at $\sqrt{s}=1$ TeV,
where $\Delta \sigma \equiv \int_{-0.5}^0 d \cos\theta 
(d\sigma / d\cos\theta)$.
This observable clearly discriminates the $\lambda=-1$ case 
from the $\lambda=+1$ case,
as its value is different by a few orders of magnitude.

\vskip 1.0cm
\noindent
\begin{center}
\begin{tabular}{|c|ccc|}\hline
{} &~~ SM~~ &~~ SM $+$ QG~~ &~~ SM $+$ QG~~  \\
  & &$(\lambda=+1)$ &  $(\lambda=-1)$ \\ \hline
~${\Delta \sigma(e^-_Le^+_R\rightarrow t_\downarrow \bar{t}_\uparrow)}/
{\Delta \sigma(e^-_Le^+_R\rightarrow t_\uparrow \bar{t}_\downarrow)}$~~ &
0.14 & 0.035 & 2.15 \\
~${\Delta \sigma(e^-_Re^+_L\rightarrow t_\uparrow \bar{t}_\downarrow)}/
{\Delta \sigma(e^-_Re^+_L\rightarrow t_\downarrow \bar{t}_\uparrow)} $~~
& 0.083 & 0.034 & 26.92 \\
\hline
\end{tabular}
\end{center}
\vspace{0.5cm}
{Table~2}. {\it The ratios of the cross sections for the next-dominant processes
to those for the dominant ones
with the angular cut $-0.5 < \cos\theta < 0$
at $\sqrt{s}=1$ TeV.
}
\vskip 0.5cm

In conclusion, we have studied the spin configuration
of the top quark pair at the process $e^+ e^- \to t \bar{t}$
with large extra dimensions.
The TeV scale quantum gravity affects
the process through the $s$-channel Feynman diagram
mediated by the Kaluza-Klein tower of spin two gravitons.
The framework of the off-diagonal basis, in which
the cross sections of the like-spin states of the top quark pair
vanish, leads to the SM result that the left-handed (right-handed) 
electron beam almost completely prefer
the spin configuration of the top and anti-top spins as Up-Down (Down-Up).
The presence of large extra dimensions modifies these features
significantly at high energies.
It yields non-zero cross sections for
the like-spin states 
and relieves the partiality of the beam polarization for a specific
spin configuration of the top quark pair.
In addition, it is shown that the angular cut $-0.5 <\cos\theta<0$ is very effective to determine the sign of the quantum gravity corrections.

\acknowledgments

We would like to appreciate valuable discussions with S.Y. Choi.
This work is supported by the Korean Science and Engineering Foundation 
(KOSEF) through the SRC program of the Center for Theoretical Physics (CTP)
at Seoul National University.

\begin{center}
\begin{figure}[htb]
\hbox to\textwidth{\hss\epsfig{file=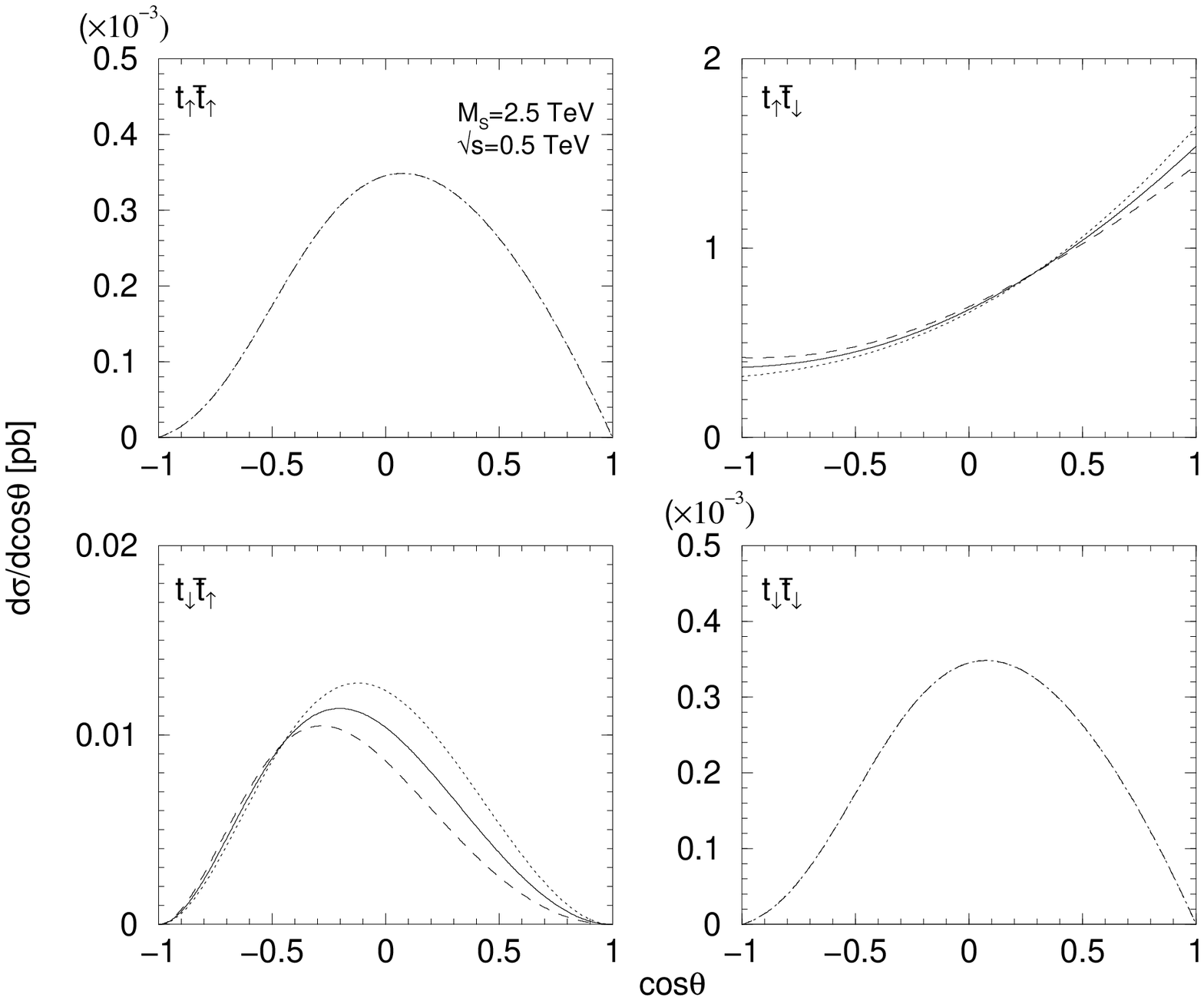,height=15cm}\hss}
\vspace{1cm}
\caption{\it The differential cross section with respect
                to the scattering angle of the top quark 
		at  $\sqrt{s}=500$ GeV with the left-handed electron beam,
                broken down to the spin configuration of the top quark pair.
The dotted (dashed) line includes the large extra dimension 
effects when $\lambda=+1$ $(\lambda=-1)$
and $M_S=2.5$ TeV.
The solid line denotes the SM background.
}
\end{figure}
\end{center}

\newpage
\mbox{ }

\begin{center}
\begin{figure}[htb]
\hbox to\textwidth{\hss\epsfig{file=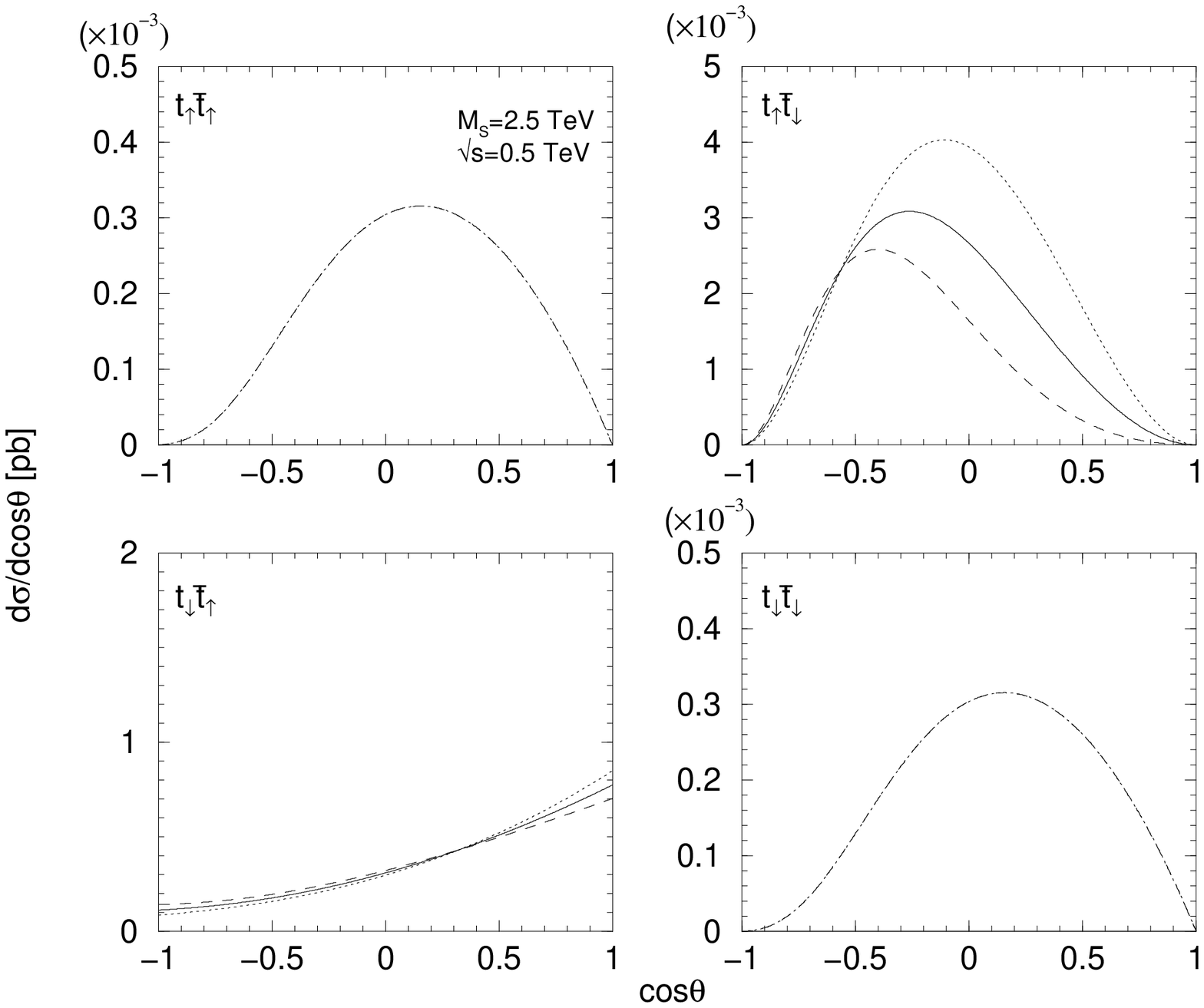,height=15cm}\hss}
% \vskip -1.5cm
\vspace{1cm}
\caption{\it The differential cross section with respect
                to the scattering angle of the top quark 
                at  $\sqrt{s}=500$ GeV with the right-handed electron beam,
                broken down to the spin configuration of the top quark pair.
The dotted (dashed) line includes the large extra dimension effects 
when $\lambda=+1$ $(\lambda=-1)$
and $M_S=2.5$ TeV.
The solid line denotes the SM background.
}
\end{figure}
\end{center}
 
\newpage
\mbox{ }

\begin{center}
\begin{figure}[htb]
\hbox to\textwidth{\hss\epsfig{file=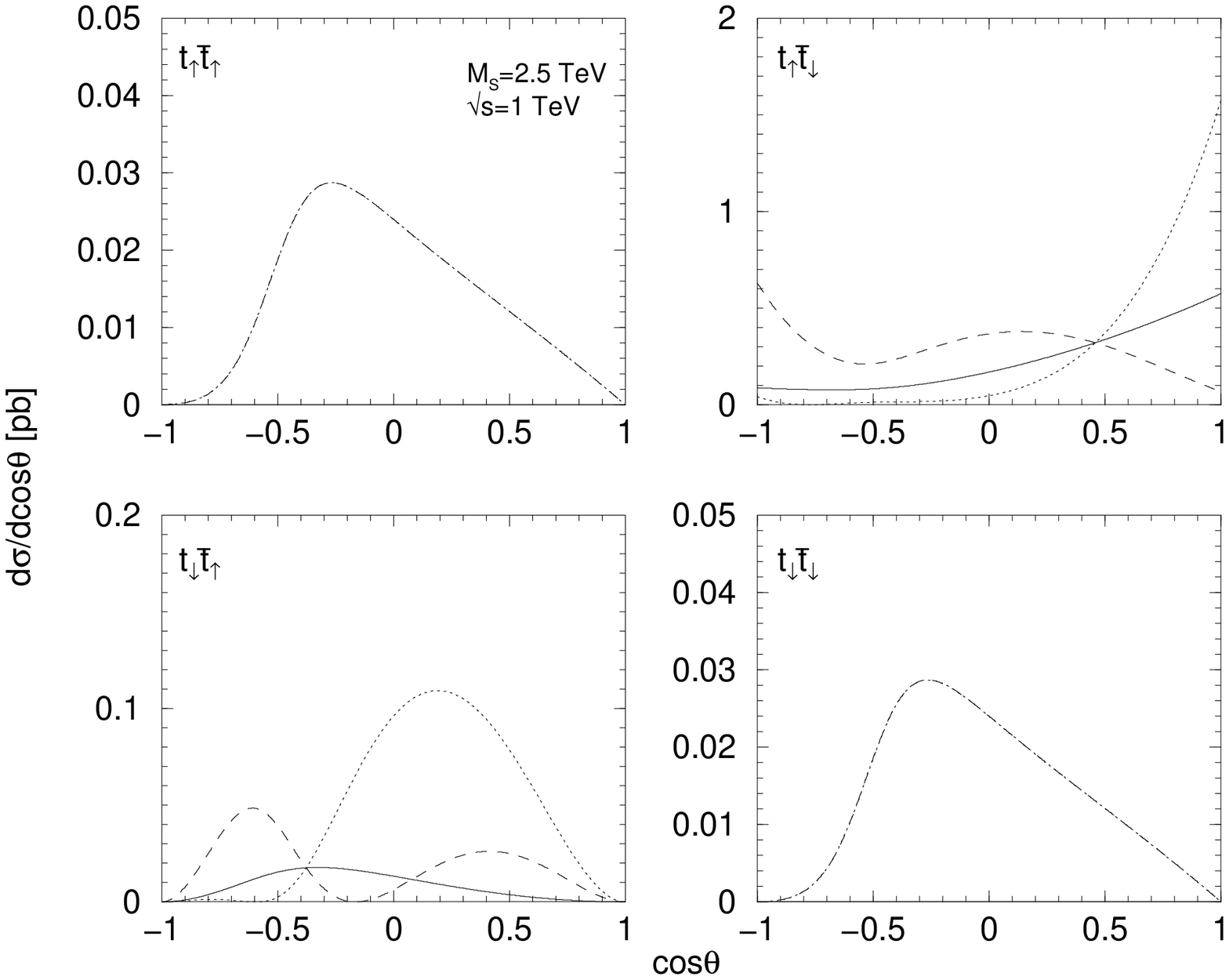,height=15cm}\hss}
% \vskip -1.5cm
\vspace{1cm}
\caption{\it The differential cross section with respect
                to the scattering angle of the top quark 
                at  $\sqrt{s}=1$ TeV with the left-handed electron beam,
                broken down to the spin configuration of the top quark pair.
The dotted (dashed) line includes the large extra dimension 
effects when $\lambda=+1$ $(\lambda=-1)$
and $M_S=2.5$ TeV.
The solid line denotes the SM background.
}
\end{figure}
\end{center}

\newpage
\mbox{ }

\begin{center}
\begin{figure}[htb]
\hbox to\textwidth{\hss\epsfig{file=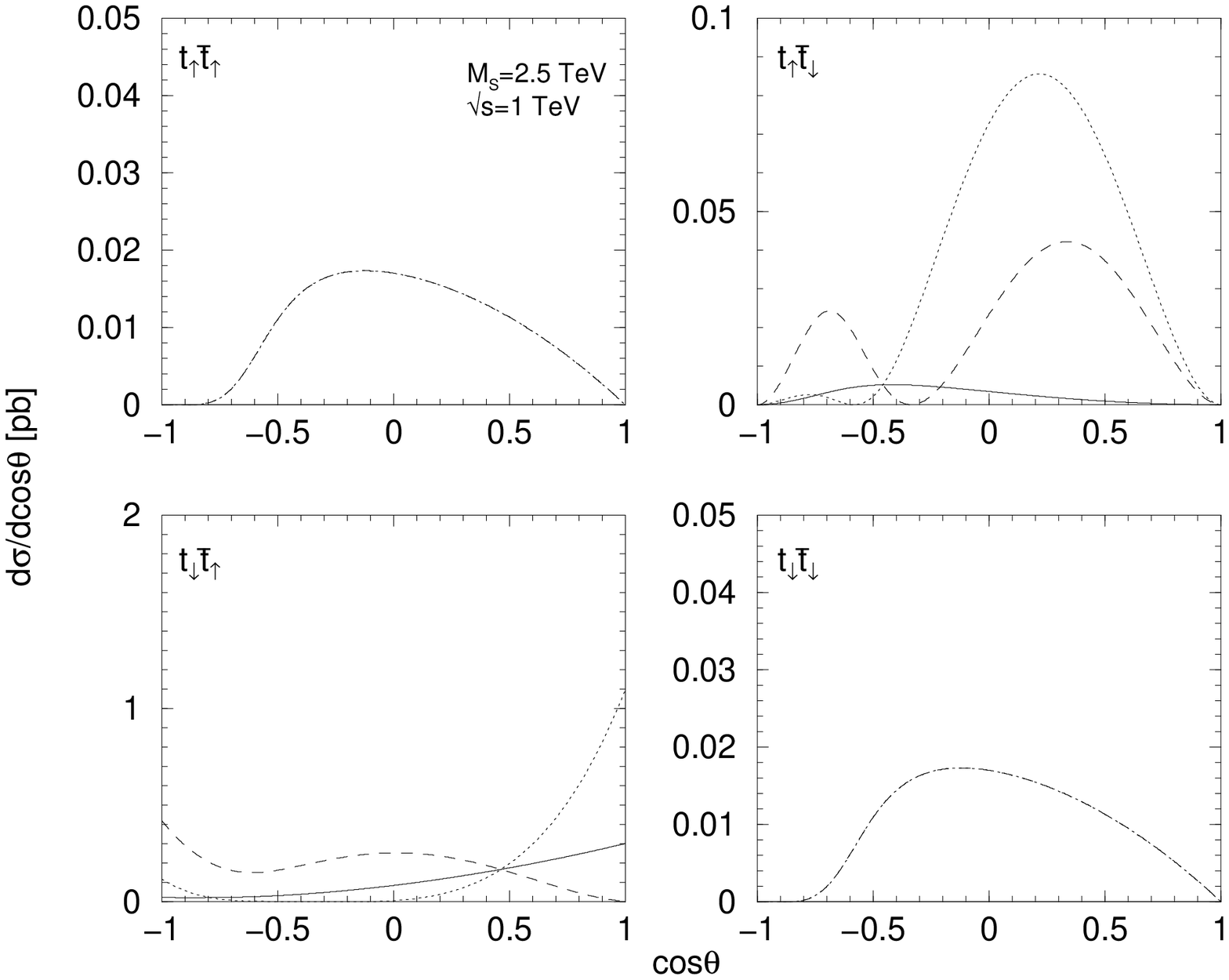,height=15cm}\hss}
% \vskip -1.5cm
\vspace{1cm}
\caption{\it The differential cross section with respect
                to the scattering angle of the top quark 
                at  $\sqrt{s}=1$ TeV with the right-handed electron beam,
                broken down to the spin configuration of the top quark pair.
The dotted (dashed) line includes the large extra dimension effects 
when $\lambda=+1$ $(\lambda=-1)$
and $M_S=2.5$ TeV.
The solid line denotes the SM background.
}
\end{figure}

\end{center}

\end{document}